\documentclass[epj]{svjour}
\usepackage{latexsym}

\usepackage{graphics}
\newcommand{\snn}{${\sqrt{s_{_{NN}}}}$}

\newcommand{\pbar}{${\rm p(\bar{p})+p}$}

\newcommand{\eea}{\end{eqnarray}}

\newcommand{\bea}{\begin{eqnarray}}

\newcommand{\nch}{N_{ch}}

\newcommand{\nnsim}{${\langle\rm N^{n}_{part}\rangle}$ }

\newcommand{\nqsim}{${\langle\rm N^{q}_{part}\rangle}$ }

\newcommand{\etaone}{|\eta| < 1}
\newcommand{\dndeta}{d\nch/d\eta}

\newcommand{\dndetaone}{${\dndeta|_{\etaone}}$}

\begin{document}
\title{Charged Particle Multiplicities in A+A and p+p Collisions 
in the Constituent Quarks Framework
}
\author{\large Rachid Nouicer}
\offprints{rachid.nouicer@bnl.gov}    
\institute{\large
Brookhaven National Laboratory, Upton, New York 11973-5000, U.S.A.
}
\date{Received: date / Revised version: date}
\abstract{Charged particle multiplicities in A+A and \pbar\ collisions
as a function of pseudorapidity, centrality and energy are studied in
both the nucleon and the constituent quark frameworks. In the present
work, the calculation of the number of nucleon and constituent quark
participants using the nuclear overlap model takes into account the
fact that for the peripheral A+A and p+p collisions can not be smaller
than two. A striking agreement is seen between the particle density in
A+A and \pbar\ collisions, both at mid-rapidity and in the
fragmentation regions, when normalized to the number of participating
constituent quarks.
The observations presented in this paper
imply that the number of constituent quark pairs participating in the
collision controls the particle production. One may therefore
conjecture that the initial states in A+A and \pbar~collisions are
similar when the partonic considerations are used in normalization.
} 
\authorrunning{R. Nouicer} \titlerunning {Charged
Hadrons Production in A+A and p+p Collisions in the Constituent Quarks
Framework} \maketitle
\section{-Introduction}
\label{intro}
Quantum Chromo-Dynamics (QCD) is the theory of strong interactions.
From a theorist's point of view strong interaction means that all
dimensionless quantities of the theory are of the order of unity.
Meanwhile, there are at least two fundamental facts about strong
interactions showing that actually it is not exactly so: there are
still certain small parameters hidden. One is that nuclei can be
viewed as made of weakly bound nucleons -- in a true strong interactions
case they would rather looke like a quark soup. The second puzzle is
that the nucleon itself can be viewed as built of three constituent
quarks (objects quasi-free non relativistic) -- in a true strong
interactions case it would rather look like a pack of an indefinite
number of quarks, antiquarks and gluons. The presence of just three
discrete objects in a nucleon is non-trivial from the point of view of the
parton model, which permits any large number of point-like
objects, i.e. quark-partons, inside a fast moving nucleon. The two
notions can be reconciled if one accepts a nucleon with two radii: a
nucleon radius $R_{n}$ that determines a mean distance between the
constituent quarks, and a proper radius of the constituent,$r_{q}$. In
such a pattern, a fast moving nucleon is a viewed as a system of
three clouds of partons, each containing a valence quark, a sea of the
quark-antiquark pairs, and gluons. QCD calculations support the
statment that inside a nucleon there are three objects of size of
0.1-0.3 fm~\cite{Shu2000}. Similar idea is applied in the study
of multihadron production processes in different types of collisions
in the framework of the picture based on dissipating energy of
participants~\cite{Edw2005}.

In this paper, I illustrate a modified version on how to calculate
the number of nucleon and constituent quark participants. One problem
in previous calculations~\cite{Bha2005,Nou2006} was confusion in
applying the appropriate minimum-bias \nqsim to the \pbar\
collisions. A second, more general problem in the previous calculation
which can be found in the Refs.~\cite{Bha2005,Nou2006,Erm2003} stems
from the fact that in the nuclear overlap model in peripheral
collisions the number of nucleon and constituent quark participants
(\nnsim and \nqsim) can be smaller than two -- this is because, in
peripheral limit, the overlap integral has a meaning of $1/2$ times
the probability to have \nnsim=2. In the previous
calculation~\cite{Bha2005,Nou2006,Erm2003}, the problem can be found
in the estimate of \nnsim for the peripheral A+A collisions and easily
from the calculation of the number of constituent quark participants
for p+p collisions such plotting \nqsim versus collision centrality or
impact parameter. A similar problem with the calculation of \nnsim for
d+Au collisions using an optical approch has been reported by
D. Kharzeev et al.~\cite{Dim2004}. The goal of the present paper is to
study the charged particle multiplicities in the nucleus-nucleus (A+A)
and nucleon-nucleon (\pbar) collisions in both the nucleon and the
constituent quark frameworks.

The motivations of this study are; nuclear collisions are studied at
unprecedented energies at the Relativistic Heavy Ion Collider (RHIC),
revealing several new phenomena embedded in a large amount of
high-quality data \cite{RHIC2005}. However, several aspects of the results
on charged particle multiplicities in the nucleon framework (results
scaled to the number of nucleon participants) are still not well
understood such as;
\vspace*{-0.08cm}
\begin{enumerate}
\item the charged particle densities near mid-rapidity region
in Au+Au at 200 GeV depend strongly on
the collision centrality. The densities have been plotted as per
nucleon participants pair~\cite{Bac2005} in order to disentangle pure nuclear
effects. The common explanation of the phenomena is due to the hard
processes. 
\item the charged particle density at mid-rapidity 
region from A+A collisions are substantially
higher than those of \pbar\ collisions at the same
energy~\cite{Bac2002,NouPanic}, 
\item the charged particle densities at the fragmentation 
region from A+A collisions are substantially
higher than those of \pbar\ collisions at the same energy~\cite{Nou2006},
\item the integrated total charged particle in A+A collisions as a
function of number nucleon participants is higher than \pbar\
collisions at the same energy indicating that there is no smooth
transition between A+A and nucleon-nucleon collisions\cite{dAu}.
\end{enumerate}
 The investigations on the aspects 1. and 2. have been already started
in Refs.~\cite{Bha2005,Nou2006,Erm2003} using the old version which
contains problem in the calculations of \nnsim and \nqsim for peripheral A+A
collisions and \pbar\ collisions. In this paper, I will review the
aspects 1. and 2. and I will extend the investigation to study the
aspects 3. and 4. in both the nucleon and constituent quark frameworks,
using the new version of the calculation.
\begin{figure}[]
\resizebox{0.45\textwidth}{!}{
\includegraphics{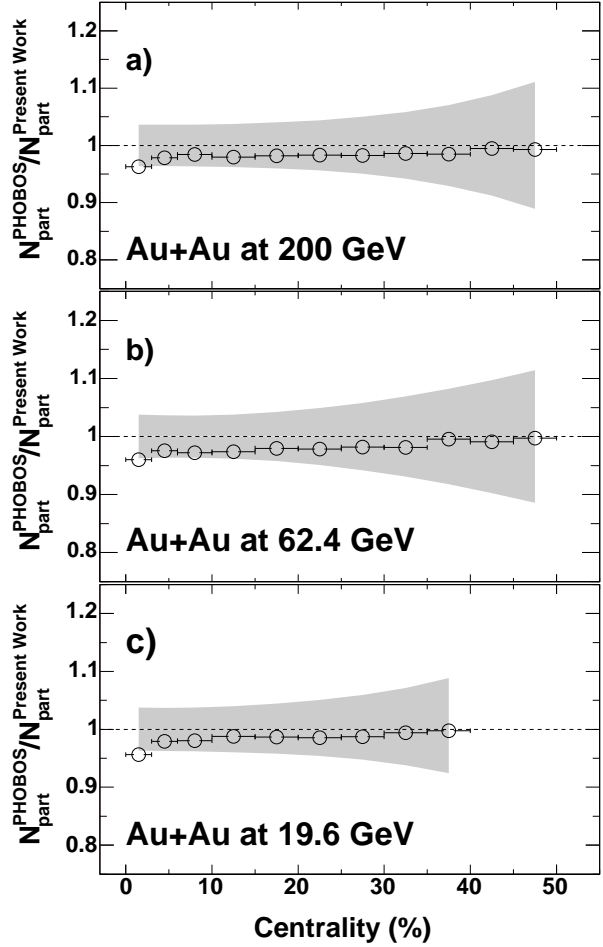} }
\caption{Quantitative evaluation of the model calculations expressed
as the ratio of the average number of nucleon participants of the
PHOBOS Glauber calculations~\cite{Bac2004} to the present work as a
function of centrality. Panel a), b) and c) correspond to Au+Au
collisions at \snn = 19.6, 62.4 and 200 GeV, respectively. The gray
bands correspond to the systematic errors on the \nnsim of PHOBOS
Glauber calculations.}
\label{fig:1}      
\end{figure}
\begin{figure}[]
\resizebox{0.5\textwidth}{!}{
\includegraphics{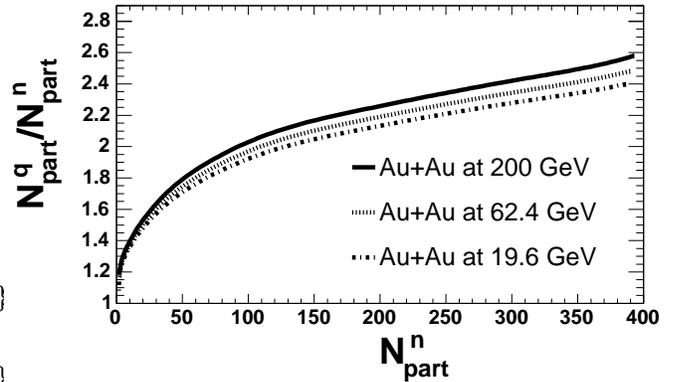}}
\caption{Ratio of \nqsim/\nnsim obtained from the present work
presented as a function of the number of nucleon participants for
Au+Au collisions at \snn = 19.6, 62.4 and 200 GeV.}
\label{fig:2}      
\end{figure}
\begin{figure}[]
\resizebox{0.5\textwidth}{!}{
\includegraphics{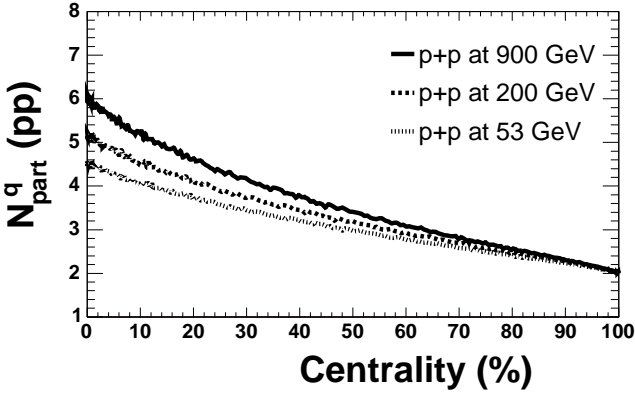}}
\caption{Distributions of constituent quark participants, \nqsim,
obtained from the present work for p+p at \snn = 53, 200, 900 GeV
represented as a function of collisions centrality.}
\label{fig:3}      
\end{figure}
\vspace*{-0.5cm}
\section{-Calculation of the Number of Participants}
\label{sec:2}
The number of nucleon participants noted by \nnsim, and the number of
constituent quark participants noted by \nqsim, are estimated using
the nuclear overlap model in a manner similar to that used in
Refs.~\cite{Bha2005,Nou2006,Erm2003} but I introduce in the present
work a modification of the calculation procedure taking into account
that for the peripheral collisions A+A and for p+p collisions in both
framewroks; the \nnsim and \nqsim can not be smaller than two. The
nuclear density profile is thus assumed to have a Woods-Saxon form,
\vspace*{-0.3cm}
\bea
        n_{A}(r) =   \frac{n_{0}}{1+\ \exp [(r-R_{n})/d]},
\eea
where $n_{0}$ is the normal nuclear density, $R_{n}$ is the nucleus
radius and $d$ is a diffuseness parameter. 

For {nucleus-nucleus} (A+B) collisions, the number of nucleon
participants, \nnsim, is in the present work calculated using the relation,
\hspace*{-0.5cm}\bea  
N^{n}_{part}|_{AB} = \int d^{2}s T_{A}(\vec{s})
P_{AB}(\vec{b})\{1-
\bigg[1-\frac{\sigma^{inel}_{NN}T_{B}(\vec{s}-\vec{b})}{B}
\bigg]^{B}\} \\ \nonumber + \int d^{2}sT_{B}(\vec{s}-\vec{b})
P_{AB}(\vec{b})\{1 -
\bigg[1-\frac{\sigma^{inel}_{NN}T_{A}(\vec{s})}{A}\bigg]^{A} \} 
\eea
where {$T(b)~=~\int_{- \infty}^{+ \infty} dz n_{A}(\sqrt{b^{2}+z^{2}})$} 
is the thickness function and
$P_{AB}(\vec{b})$ 
is defined by: 
\bea 
P_{AB}(\vec{b})=
\frac{1}{1- \exp(-\sigma^{inel}_{NN} T_{AB}(\vec{b}))} 
\eea 
where
$T_{AB}(\vec{b})$ is the overlap function defined as the
product of the thickness functions of the colliding nuclei A and B,
integrated over the two transverse dimensions: 
\bea 
T_{AB}(b)=\int d^{2}s' T_{A} (\vec{s'}) T_{B}(\vec{s'}-\vec{b}) 
\eea

$A$ and $B$ are the
mass number of the two colliding nuclei and the ${\sigma^{inel}_{NN}}$
is the inelastic nucleon-nucleon cross section. In the present work
I use the FRITIOF parameterization $R_{n} = 1.16~A^{1/3} - (1.16)^{2}~A^{-1/3}$ fm and
$d$ = 0.54 fm which are well within the measurements
of electron scattering from Au nuclei~\cite{Hah1956}.  

Fig.~\ref{fig:1} shows the ratio of \nnsim obtained by PHOBOS~\cite{Bac2004}
and to the \nnsim from the present work for Au+Au at \snn = 19.6, 62.4
and 200 GeV. The \nnsim and \nqsim from the present work are presented
on Table~\ref{tab:1} for Au+Au collisions at RHIC energies.

The number of constituent quark participants, \nqsim, is calculated in
a similar manner by taking into account the following changes related
to the physical realities and also taking into account that in
peripheral collisions the number of \nqsim can not be smaller than
two:
\begin{enumerate}
\item the density is three times that of nucleon
density with $n_{0}^{q}$ = 3$n_{0}$ = 0.51 ${\rm fm^{-3}}$,
\item the cross sections $\sigma_{CQ}$ = $\sigma^{inel}_{NN}$/9,
\item the mass numbers of the colliding nuclei are three times their
values, keeping the size of the nuclei same as in the case of \nnsim.
\end{enumerate}

Fig.~\ref{fig:2} presents the ratio of \nqsim/\nnsim\ as a function of
\nnsim\ for Au+Au collisions at RHIC energies. The ratio shows that
the correlation between \nnsim\ and \nqsim\ is not linear and that it
depends on the colliding energy.

For proton-proton (p+p) collisions the same procedure has been used to
calculate the number of constituent quark participants by using $A$ =
3 and $B$ = 3 and the nuclear density profile is assumed to have a
sharp sphere form with uniform radii of 0.8 fm~\cite{Won1994}. The
reason to use sharp sphere density profile is because the Wood-Saxon
density profile becomes unrealistic for low A. The systematic error
related to this approximation can be estimated by running the code with B=1 
(for p+A), and comparing obtained TAB(0) value with TA(0) (for
impact parameter b=0). The result is presented in Table~\ref{tab:2}. The
discrepancy is larger for Wood-Saxon profile with low A. The \nqsim\
for central, 0-6\%, and minimum bias
of p+p collisions are presented on the
Table~\ref{tab:3}. Fig.~\ref{fig:3} shows the distributions of the
number of constituent quark participants obtained from the present
work elucidated as a function of collisions centrality for
nucleon-nucleon (p+p) collisions at \snn= 53, 200 and 900 GeV.
\begin{table}
\caption{Numbers of constituent quark participants,\nqsim,
obtained from the present work elucidated as a function of collision
centrality in Au+Au collisions at \snn= 19.6, 62.4 and 200 GeV.
\label{tab:1}
}
\hspace*{0.5cm}\begin{tabular}{|c|c|c|c|}
\hline
\multicolumn{1}{|c|}{Centrality (\%)}&
\multicolumn{1}{|c|}{200 GeV}&
\multicolumn{1}{|c|}{62.4 GeV}&
\multicolumn{1}{|c|}{19.6 GeV}\\
\hline
Bin&\nqsim &\nqsim & \nqsim\\ 
\hline
0-3 &  952.5& 
       907.2&       
       869.4\\
\hline 
3-6 &  837.7& 
       796.0&
       761.4\\
\hline 
6-10 &  731.4& 
        693.1&
        661.8\\
\hline 
10-15 & 614.5& 
        580.4&
        552.8\\
\hline 
15-20 & 501.7& 
        472.1&
        448.4\\
\hline 
20-25 &  408.5&
         382.9&
         362.7\\
\hline 
25-30 &  331.6& 
         309.7&
         292.6\\
\hline 
30-35 &  265.3& 
         246.8&
         232.4\\
\hline 
35-40 &  209.1& 
         193.5&
         181.6\\
\hline 
40-45 &  161.6& 
         148.8&
         139.1\\
\hline 
45-50 &  123.1& 
         112.8&
         105.1\\
\hline
\end{tabular}
\end{table}
\begin{center}
\begin{table}
\caption{Values of overlap function TAB and TA for impact parameter
b=0 calculated using densities profiles Wood-Saxon and sharp sphere 
for p(A=1)+B collisions using the nuclear overlap model}
\label{tab:2}       
\hspace*{1.3cm}
\vspace*{-0.3cm}
\begin{tabular}{|c|c|c|c|}
\hline
profile & A & B & TAB(0)/TA(0) \\
\hline
Wood-Saxon & 1 & 236 & 0.9  \\
Wood-Saxon & 1 & 12 & 0.6  \\
\hline
sharp sphere & 1 & 236 & 1.0  \\
sharp sphere & 1 & 12 & 0.9  \\
\hline
\end{tabular}
\end{table}
\end{center}
\begin{table}
\caption{Values of \nqsim obtained from the present work for most
central and minimum-bias (min$-$bias) of \pbar\ collisions at several
collisions energies.}
\label{tab:3}
\vspace*{-0.1cm}
\hspace*{-0.2cm}
\begin{tabular}{|c|c|c|c|c|c|c|}
\hline
${\rm \sqrt{s_{_{NN}}}}$ (GeV) & 53 & 200 & 540 & 630 & 900 & 1800\\
\hline
${\sigma_{CQ} = \sigma_{NN}/9}$ & 3.89 & 4.66 & 5.33 & 5.44 & 5.66 & 6.22\\
\hline
\nqsim  (0-6\%) &  4.37 &4.94     &5.46     & 5.55    & 5.72    & 5.98    \\
\hline
\nqsim (min-bias)& 3.10 & 3.29    &3.47     & 3.51    & 3.57   & 3.73    \\
\hline 
\end{tabular}
\end{table}
\begin{figure}[]
\begin{center}
\resizebox{0.45\textwidth}{!}{
\includegraphics{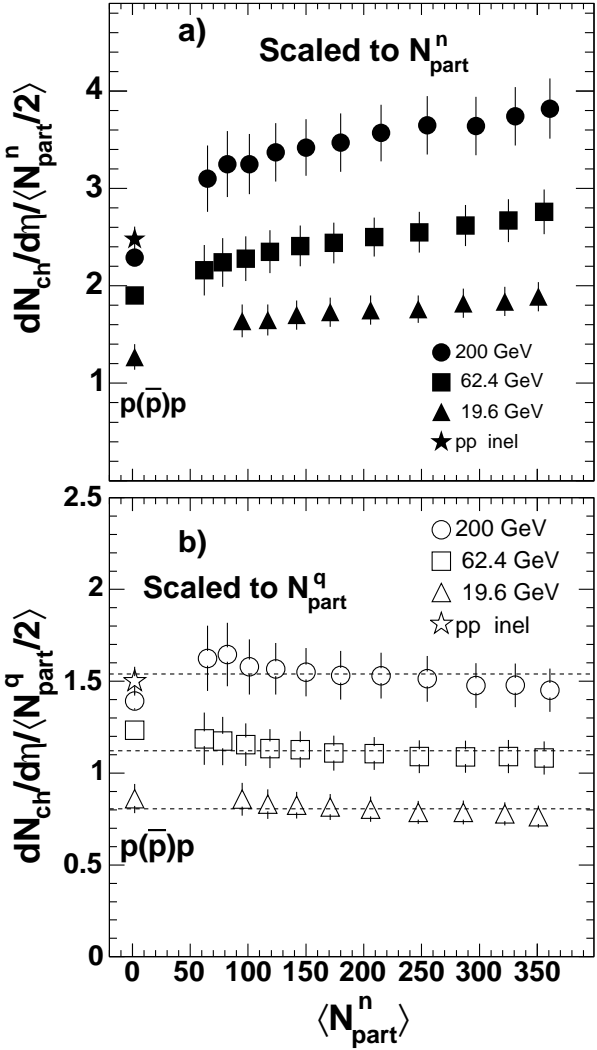}}
\caption{Particle density, \dndetaone, divided by a) the number of
nucleon participants pair, b) the number of constituent quark
participants pair in Au+Au and \pbar\ collisions at three
energies~\cite{Bac2005,Bac2004,Abe}. The dashed lines in panel b)
correspond to the average of Au+Au data at given energy. The \pbar\
for non single diffractive (NSD) data are represented with similar
symbols as Au+Au collisions at the same energy at \nnsim=2. The star
symbols in panel a) and b) correspond to \pbar\ inelastic data at \snn
= 200 GeV. The \pbar\ data in in panel b) have been normalized by the
minimum-bias \nqsim.}
\label{fig:4}      
\end{center}
\end{figure}

\vspace*{-1cm}
\section{-Physics Results and Discussions}
\label{sec:3}
 Fig.~\ref{fig:4} shows the centrality dependence of the charged
particle density per participant pair. Fig.~\ref{fig:4}.a shows the
PHOBOS results~\cite{Bac2005,Bac2004} on \dndetaone\ per nucleon
participant pair for Au+Au at 19.6, 62.4 and 200 GeV. The centrality
dependence of the mid-rapidity yields has often been interpreted in a
two component picture of particle production by soft and hard
processes. As the beam energy increases, particle production from hard
processes, which exceed the number of participants pairs by a factor
$\sim$ 5-6 in central events for \snn\ ranging from 19.6 to 200 GeV,
is expected to dominate over that from soft processes as the
mini-jets cross sections increase~\cite{Gyu1994}. Fig.~\ref{fig:4}.b
shows the centrality dependence of \dndetaone\ per constituent quark
participants pair. I observe a constant or a slightly decreasing dependence of
($dN/d\eta$)/(\nqsim/2) on centrality from peripheral to central collisions for Au+Au
at 200 GeV and good scaling of ($dN/d\eta$)/(\nqsim/2) on centrality for
Au+Au at 19.6 and 62.4 GeV. In Fig.~\ref{fig:4}.b, I extend
the study presented in Ref.~\cite{Erm2003} from Au+Au at 200 GeV to
19.6 and 62.4 GeV using the new calculations. I agree with
the interpretation presented in Ref.~\cite{Erm2003} that the
experimentally observed increase of (\dndetaone)/(\nnsim/2) can be
explained by the relative increase in the number of interacting
constituent quarks in more central collisions. In the present work, I
add to this study \dndetaone\ of \pbar\ inelastic as well NSD
collisions. In Fig.~\ref{fig:4}.b, I observe that the \pbar\ data
are in good agreement with Au+Au data for different collision
centrality at the same energy.  It should be noted that in
Fig.~\ref{fig:4}.b the \dndetaone\ of ${\rm p(\bar{p})+p}$ data (open
symbols) have been normalized by minimum-bias \nqsim presented in
table~\ref{tab:3}.
\begin{figure}[]
\begin{center}
\resizebox{0.5\textwidth}{!}{
\includegraphics{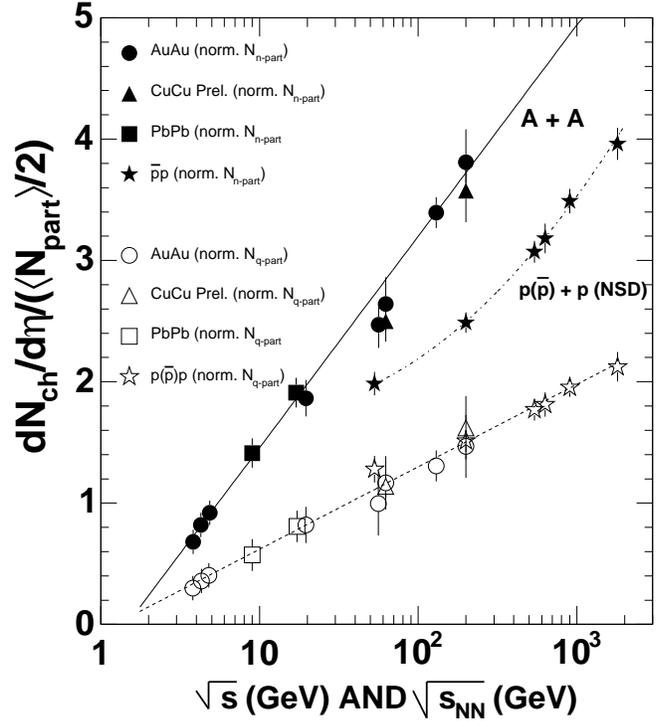}}
\caption{Particle density per constituent quark participant pair
(open symbols) and particle density per nucleon participant pair
(solid points) produced in central (0-6\%) nucleus-nucleus (A+A)
collisions presented as a function of collision energy at AGS,
SPS~\cite{Ahl,Bach,Blu} and RHIC~\cite{Gunt} and in ${\rm
p(\bar{p})+p}$ collisions (inelastic)\cite{Abe}. The errors bars
correspond to the systematic errors. The solid line represents a
linear fit through solid symbols for A+A data, ${\rm f_{AA} = -0.287 +
0.757 ln (\sqrt{s}) }$. The dashed~dotted~line corresponds to the fit
through solid symbols of ${\rm p(\bar{p})+p}$ collisions, ${\rm f_{pp}
= 2.25-0.41 ln(\sqrt{s})+ 0.09 ln^{2}(\sqrt{s})}$.  The dashed line
corresponds to a linear fit trough the open symbols in the constituent
quarks framework, ${\rm f_{ p(\bar{p})p/AA} =
-0.06+0.3ln(\sqrt{s})}$. It should be noted that the \dndetaone of
${\rm p(\bar{p})+p}$ data (open symbols) have been normalized by
minimum-bias \nqsim presented in table~\ref{tab:3}.}
\label{fig:5}      
\end{center}
\end{figure}  

\begin{figure}[]
\resizebox{0.45\textwidth}{!}{
\includegraphics{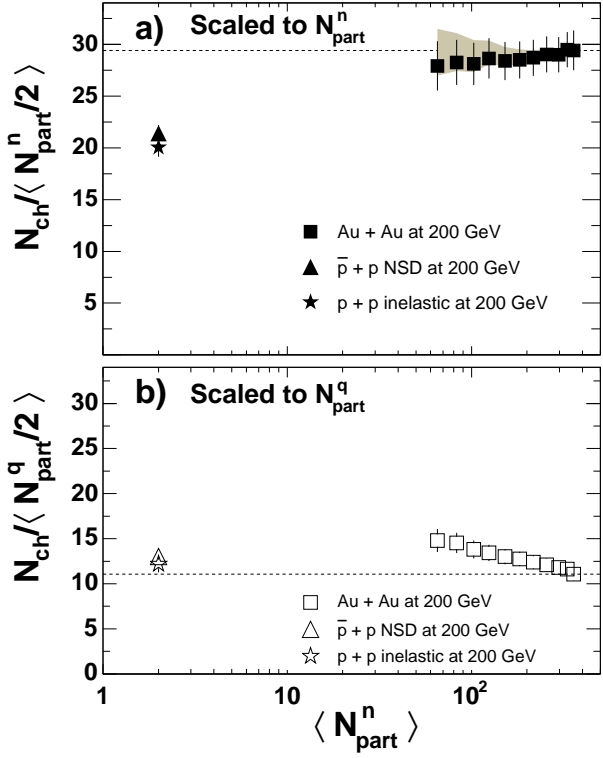}}
\caption{Integrated total charged particles obtained from Au+Au
collisions and \pbar\ inelastic as well NSD collisions at the same
energy \snn= 200 GeV \cite{Bac2005,Bac2004,Abe} normalized by a) number of nucleon participant
pair, b) number of constituent quarks participant pair. The errors
shown with vertical bars are full systematic errors. In panel b) the
N$_{ch}$ for \pbar\ data (open symbols) have been normalized by
minimum-bias \nqsim. The dashed lines
are guide lines relatively to the most central collisions in Au+Au
system}
\label{fig:6}      
\end{figure}

Fig.~\ref{fig:5} shows the primary charged particle density for
central collisions at mid-rapidity divided by a) the number of
participant nucleon pairs (\nnsim/2) as solid symbols and b) the
number of participant constituent quark pairs (\nqsim/2) as open
symbols presented as a function of collision energy. The data
are for Au+Au collisions at AGS, Pb+Pb collisions at the
CERN-SPS~\cite{Ahl,Bach,Blu} and for Au+Au and Cu+Cu collisions at
RHIC~\cite{Gunt}. Also shown for comparison are results from p(${\rm
\bar{p}}$)+p collisions~\cite{Abe}. 

In the nucleon participants
framework (solid symbols in Fig~\ref{fig:5}), the particle density per
nucleon participant pair for A+A collisions (solid points) shows an
approximately logarithmic rise with $\sqrt{s_{_{NN}}}$ over the full
range of collision energies.  The comparison of the particle density
per nucleon of Au+Au to Cu+Cu collisions at the same energies, ${\rm
\sqrt{s_{NN}}}$ = 62.4 and 200 GeV indicates that in symmetric
nucleus-nucleus collisions the density per nucleon participant does
not depend on the size of the two colliding nuclei but only on the
collision energy. This means that for Si+Si collisions at ${\rm
\sqrt{s_{NN}}}$ = 200 GeV, the particle density per nucleon
participant will be similar to Au+Au collisions at the same
energy. Also I observe that the charged particle multiplicity per
participant nucleon pair (solid symbols) in A+A collisions is higher
compared to \pbar\ collisions at the same energy. It should be noted
that number of nucleon participants for \pbar\ has been chosen to be 2
(\nnsim (pp) =2).

In contrast, I observe that the particle density per constituent
quark participant pair (open symbols in Fig~\ref{fig:5}) is similar
for nucleus-nucleus collisions and nucleon-nucleon collisions at the
same energy. It thus appears that using partonic participants accounts
for the observed multiplicity in both A+A and \pbar\ collisions. One
may therefore conjecture that the initial states in A+A and \pbar\
collisions are similar. It should be noted that Fig~\ref{fig:5} is
similar to Fig.6 presented in Ref.~\cite{Bha2005} but it was done
independently, extended to the Cu+Cu system, and used the new version of
calculation presented in this paper.

Fig.~\ref{fig:6} shows the integrated total charged particle divided
by a) per nucleon participant pair ($N_{ch}$/(\nnsim/2)), b) per
constituent quarks participant pair ($N_{ch}$/(\nqsim/2)) for
Au+Au and \pbar\ collisions at the same energy, 200 GeV. In the nucleon
participant framework, panel a), the integrated total charged particle
in Au+Au collisions, as a function of number nucleon participants, is
higher than p(${\rm \bar{p}}$)+p collisions indicating that there is
no smooth transition between A+A and nucleon-nucleon
collisions~\cite{dAu}. This observation can be strongly confirmed by
adding the preliminary data from Cu+Cu presented by PHOBOS
collaboration~\cite{Gunt} (not shown in this figure) which correspond to the
peripheral data of Au+Au system at the same energy. These comparisons
between A+A and \pbar\ collisions are, however, based on scaling with the
number of nucleon participants. In the constituent quarks
framework, panel b), the $N_{ch}$/(\nqsim/2) of p+p NSD (or
inelastic) collisions agree with the central Au+Au collisions at the same energy. 
\begin{figure}[]
\resizebox{0.45\textwidth}{!}{
\includegraphics{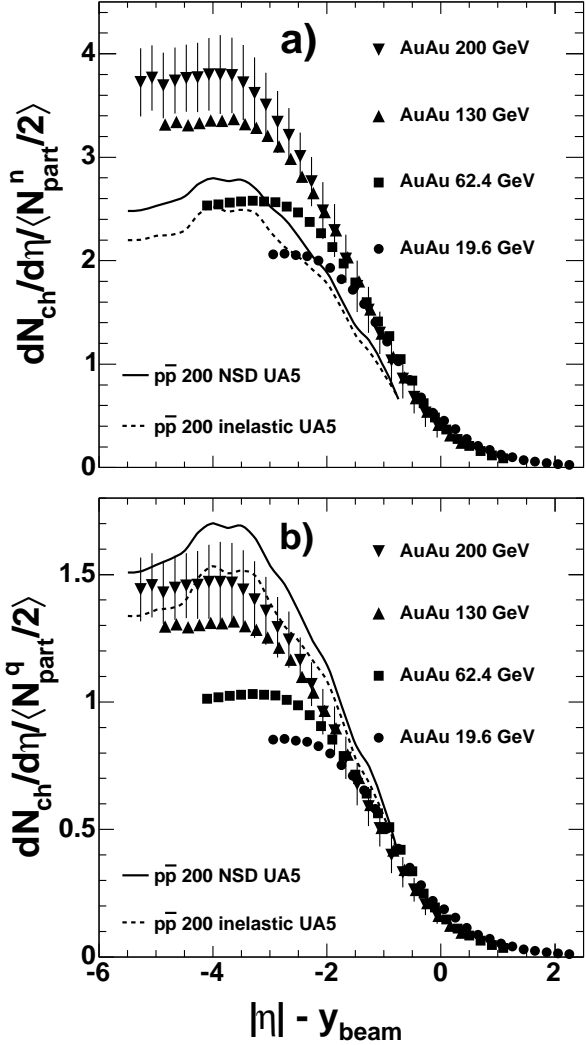}}
\caption{Pseudorapidity distributions of charged particle for Au+Au
   collisions at RHIC energies~\cite{Bac2005} and p(${\rm \bar{p}}$)+p collisions at
   200 GeV~\cite{Abe}. The distributions have been shifted to ${\rm
   \eta -y_{beam}}$ in order to study the fragmentation regions in one
   of the nucleus rest frame.  Panels a) and b) correspond to
   $dN_{ch}/d\eta$ distributions scaled to the number on nucleon
   participants pair and to the number of constituent quark
   participants pair, respectively. The systematic errors have been
   shown just for Au+Au at 200 GeV for clarity. In panel b) the \pbar\
   data have been normalized by the minimum-bias \nqsim.}
\label{fig:7}      
\end{figure}

In general, the charged particle production in the 
fragmentation region is thought to be distinct from that at
mid-rapidity, although there is no obvious evidence for two separate
regions at any of the RHIC energies. This observation is made based on
the $dN_{ch}/d\eta$ distributions of charged particle presented in
Ref.~\cite{NouPanic,Mark1}. Fig.~\ref{fig:7} shows the charged
particles produced in the fragmentation region for the most
central (0-6\%) Au+Au collisions at four RHIC energies compared to \pbar\
(inelastic and (NSD)) collisions at 200 GeV.  When normalized to
\nnsim/2, Fig.~\ref{fig:7}.a, I observe that the multiplicity in the
limiting fragmentation region in A+A collisions is higher than for
p(${\rm \bar{p}}$)+p collisions at the same energy,\snn=200 GeV.  If,
however, the comparison is carried out for multiplicities normalized
to \nqsim/2, Fig.\ref{fig:6}.b, A+A and \pbar\ collisions exhibit a
striking degree of agreement. Again, this observation implies that the
number of constituent quark pairs participating in the collision
controls the particle production in the central collisions.

\section{-Conclusions}
\label{sec:4}
The charged particle production results from A+A and \pbar\ 
collisions have been compared based on the number of
nucleon participants and the number of constituent quark 
participants. In both normalizations, I observe that the charged
particle densities in Au+Au and Cu+Cu collisions are similar for both
\snn = 62.4 and 200 GeV.  This implies that in
symmetric nucleus-nucleus collisions the charged particle density does
not depend on the size of the two colliding nuclei but only on the
collision energy. In the nucleon participants framework, the particle
density at mid-rapidity as well as in the fragmentation
region from A+A collisions are higher than those of \pbar\ 
collisions at the same energy. Also the multiplicity of total
charged particle, in A+A collisions, as a function of number nucleon
participants is higher than \pbar\ collisions at the
same energy indicating that there is no smooth transition between
peripheral A+A and nucleon-nucleon collisions.  However, when the
comparison is made in the constituent quarks framework, A+A and
\pbar\ collisions exhibit a striking degree of
agreement. The observations presented in this paper imply that the
number of constituent quark pairs participating in the collision
controls the particle production. One may therefore conjecture that
the initial states A+A and \pbar\ collisions are similar when the
partonic considerations are used in normalization. 

\begin{acknowledgement}
I would like to express my gratitude to Bhaskar De for making his code
available. This allowed us to cross check his code written in FORTRAN
to our code written in C$^{++}$; the languages of the codes are
different but the formula are the same. I would like also to join my
gratitude to S. Eremin and S. Voloshin to thank D. Miskowiec for
making his code ``nuclear ovelap model'' available. Also I would like
to thank M. D. Baker and B. Wosiek for valuable advice concerning
the present work. This work was supported by U.S. DOE
Grant No. DE-AC02-98CH10886.
\end{acknowledgement}

\end{document}